# Chromatic acclimation and population dynamics of green sulfur bacteria grown with spectrally tailored light

Semion K. Saikin,[a,†] Yadana Khin,[b] Joonsuk Huh,[a] Moataz Hannout,[c] Yaya Wang,[b] Farrokh Zare,[b] Alán Aspuru-Guzik,[a] Joseph Kuo-Hsiang Tang[b,*]

[a] Department of Chemistry and Chemical Biology, Harvard University, Cambridge, MA 02138 USA; [b] School of Chemistry and Biochemistry, Clark University, Worcester, MA 01610-1477 USA, [c] Department of Physics, Clark University, Worcester, MA 01610-1477 USA

To whom correspondence should be addressed:
[†]E-mail: saykin@fas.harvard.edu, Tel: 1-617-496-8221, Fax: 1-617-496-9411
[*]E-mail: jtang@clarku.edu, Tel: 1-614-316-7886, Fax: 1-508-793-8861

**Living organisms have to adjust to their surrounding in order to survive in stressful conditions. We study this mechanism in one of most primitive creatures – photosynthetic green sulfur bacteria. These bacteria absorb photons very efficiently using the chlorosome antenna complexes and perform photosynthesis in extreme low-light environments. How the chlorosomes in green sulfur bacteria are acclimated to the stressful light conditions, for instance, if the spectrum of light is not optimal for absorption, is unknown. Studying *Chlorobaculum tepidum* cultures with far-red to near-infrared light-emitting diodes, we found that these bacteria react to changes in energy flow by regulating the amount of light-absorbing pigments and the size of the chlorosomes. Surprisingly, our results indicate that the bacteria can survive in near-infrared lights capturing low-frequency photons by the intermediate units of the light-harvesting complex. The latter strategy may be used by the species recently found near hydrothermal vents in the Pacific Ocean.**

INTRODUCTION

Many species of photosynthetic microbes live in low-light environments, where other light harvesting organisms such as higher plants cannot survive. For example, some types of green sulfur bacteria have been found in Black Sea about 80 meters below the surface (1) and also in the deep-sea microbial mat (2), where only geothermal light is available. To support their life cycle with enough energy in such extreme ecological niches, photosynthetic microbes use light-harvesting antennae – molecular complexes that are responsible for the light absorption and energy transfer (3). The structures and the molecular compositions of light-harvesting complexes (LHCs) are both dynamically regulated, which, in turn, allows the bacteria to adjust to their specific environments. On the timescales comparable to the population doubling time, the regulation in the LHC structure occurs at the epigenetic level (4) and is considered as an acclimation.

Photosynthetic microbes exploit various acclimation strategies. For example, some species of cyanobacteria can sense the light frequency using specific photoreceptors and then dynamically adjust the structure of antenna complexes correspondingly (5,6). Purple bacteria respond to low-light conditions by changing the ratio and the structure of the LH1 and LH2 complexes (7,8). It has also been reported that green photosynthetic bacteria modify the molecular composition of the light-harvesting complexes depending on light intensities (9,10).

In this study, we focus on how green sulfur bacteria respond to the spectral distribution of available light. These bacteria are obligate photoautotrophs (11), and they survive in extreme low-light environments (1,2). Living in anoxygenic conditions, green sulfur bacteria are assumed to be one of the first organisms on Earth that started using solar light to drive intracellular metabolic reactions. Presently, they colonize several types of environments such as sulfur hot springs, lakes, and the Black Sea, to name a few (11). Thus, the study of their response to spectral properties of the light available may provide us with a better understanding of the origins of phototrophic life on Earth, and also with new ideas for the design of smart artificial solar light collectors.

Surprisingly, despite the fact that the amount of energy available for photosynthesis is determined by the spectral distribution of light, most of available studies on acclimation of LHC in green sulfur bacteria are focused on the light intensity (10,12-17). To address this problem we analyze the absorption properties and population dynamics of the *Chlorobaculum* [*Cba.*] *tepidum* species of green sulfur bacteria grown with light-emitting diodes (LEDs) that provide spectrally tailored light in varied frequencies and intensities.

Green sulfur bacteria are one of two known bacterial families that use chlorosomes in the LHC. (Another family of photosynthetic bacteria producing chlorosomes is filamentous anoxygenic phototrophs.) The chlorosome is an organelle, which is mostly composed of bacteriochlorophyll (BChl) pigment molecules self-assembled in a multilayer structure, **Fig. 1**, (18,19). It is generally accepted that low-light adaptation of green bacteria is owed to the presence of the chlorosome, which has a large absorption cross section and can capture photons very efficiently. Two other components of LHC are the baseplate (20) and the Fenna-Matthews-Olson (FMO) protein complex (21), where the BChl molecules embedded in proteins. The energy of photons absorbed by the chlorosome is transmitted through these units consequently to the reaction center as molecular excitations. In *Cba. tepidum* the chlorosome is composed of BChl *c* molecules, while the baseplate and the FMO complex contain BChl *a* pigments. The $Q_y$ absorbance peaks for the chlorosome, baseplate complex, FMO protein and reaction center are about 750 nm, 792 nm, 810 nm and 840 nm, respectively, see **Fig. 1**.

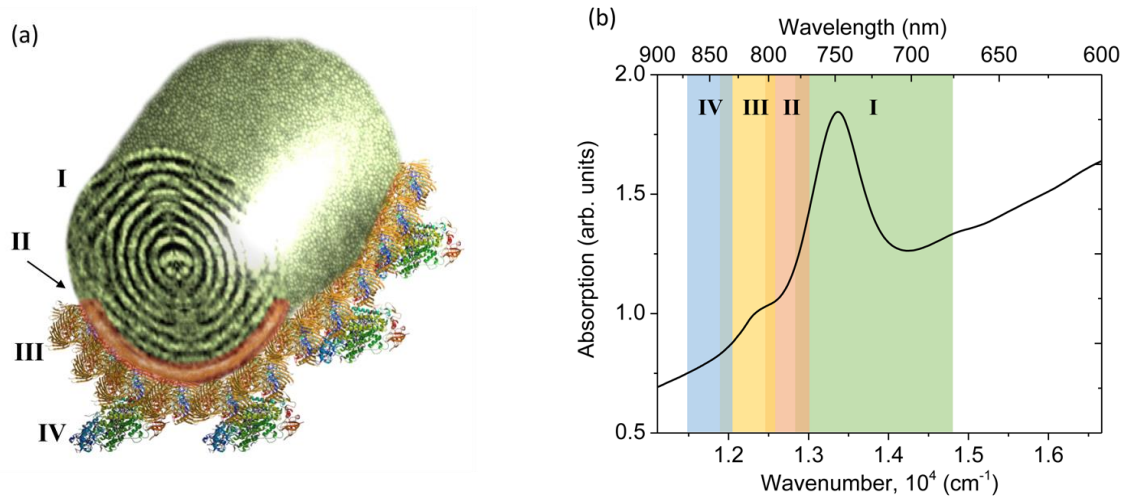

**Figure 1.** (a) Schematic structure of the light-harvesting complex (LHC) in green sulfur bacteria: I –chlorosome; II – baseplate; III – Fenna-Matthews-Olson protein complexes; IV – reaction centers. (b) $Q_y$-band absorption of a bacterial culture, the marked ranges correspond to different structural units of LHC. In *Cba. tepidum* the structural unit I is composed of BChl *c* pigments, while the units II-IV contain BChl *a* pigments.

RESULTS

Bacterial cultures were grown for 46 to 140 hours, depending on the light source, at what is considered the optimal growth temperature (22), $T = 47 \pm 3$ ºC. To illuminate cultures we used seven different light sources: 700, 750, 780, 800, 850, and 940 nm LEDs and also 850 nm LED with a band-pass filter, which narrows the LED emission spectrum to 10 nm. While the actual spectra of LEDs are broad and the peak values are slightly shifted from the nomenclature used by the companies we still use it. The population dynamics was monitored using culture optical density at the 625 nm wavelength (23), where culture absorption is minimal and optical density mostly determined by the light scattering from bacteria. It appears that the cultures can grow with all light sources except that of the 940 nm LED, which is far below the absorption transition of the reaction centers.

It is interesting to note that bacterial cultures can grow to normal populations illuminated with a narrow band 850 nm light, which is approximately 100 nm red-shifted from the $Q_y$ absorption maximum of the chlorosome. In contrast, bacterial cultures illuminated with 700 nm LEDs were found to be barely alive. The population dynamics for the 700 nm cultures was qualitatively different, as compared to other five cultures, and we did not calculate the growth rates for these. For the rest of the cultures, population dynamics can be described using a conventional sigmoidal curve as is illustrated in **Fig. 2** for bacteria grown with 850 nm LEDs. The data points were fitted using the relation

$$n_t = \frac{k}{\frac{k-n_0}{n_0}e^{-rt}+1} \quad , \tag{1}$$

where $k$ is the saturation density, $n_0$ is the initial density and $r$ is the growth rate, which is dependent on the light intensity. The obtained growth rates, see Table I, are comparable for 750, 780, and 800 nm LEDs, while the cultures grown with 850 and 850 nm band-pass filter LEDs grow more slowly.

**Table I.** Power of light-emitting diodes together with the growth rates of bacterial cultures.

| LEDs | Growth rate (hr$^{-1}$) | LED power (mW) |
|---|---|---|
| 700 nm | – | 5 |
| 750 nm | 0.28±0.03 | 150 |
| 780 nm | 0.23±0.03 | 450 |
| 800 nm | 0.29±0.03 | 450 |
| 850 nm | 0.11±0.01 | 200 |
| 850 nm filter | 0.05±0.01 | 40 |

The population dynamics of cultures grown with 700 nm LEDs is illustrated in **Fig. 2**. While these cultures where developing initially, on 12-24 hours timescales their population saturated at about 1/10 of the population of healthy cultures. After 3 days of growth, the 700 nm cultures where barely alive. Additionally, at the saturation level all healthy cultures were of green color, while the color of 700 nm culture was yellow-green (see SI for images).

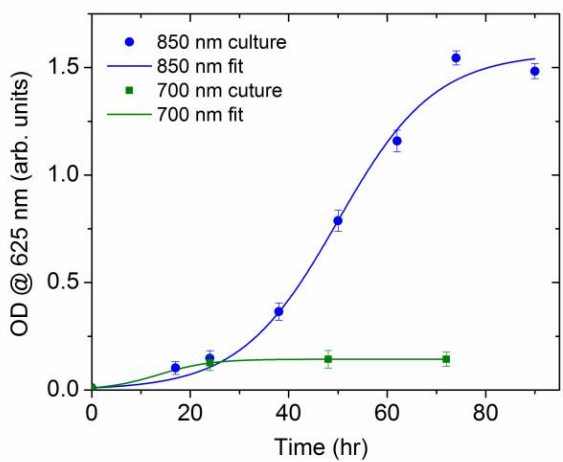

**Figure 2.** Population dynamics of bacterial cultures illuminated with 850 nm and 700 nm LEDs and monitored by optical density at 625 nm. The experimental data for a single set (6 replications) of experiments are fitted with sigmoidal curves. The error bar corresponds to standard error of mean.

**Figure 3** shows cell images for the *Cba. tepidum* cultures irradiated with 750 nm LEDs and 850 nm LEDs with a band-pass filter. Cells irradiated with 750 nm LED exhibited images similar to ones reported by Bryant and co-workers, whereas 850 nm cultures grown exhibited more long-rod and/or aggregated cells that may result from growing under light-stress conditions (24).

Extinction spectra of the cultures in the wavelength range of 650 – 900 nm are mostly determined by the $Q_y$-band absorption of the BChl pigments in the LHC, see **Fig. 1(b)**. We use this feature to analyze acclimation of the LHC in bacterial cultures to spectral composition of light. In order to compare dynamics of LHCs the measured OD spectra were normalized to the extinction at 625 nm. Then, the baseline, which characterizes off-resonance light scattering, was removed. The resulting spectra measured after at least two days of culture growth together with the emission spectra of LEDs are shown in **Fig 4**. At this timescale, the peak intensities saturate to stationary values as it is shown for the 750 nm culture. The main peak about 750 nm and the short wavelength wing correspond to the chlorosome (BChl *c*) absorption, while the long wavelength band at about 800 nm is due to the BChl *a* pigments from FMO and the baseplate. For all cultures the spectra of LEDs overlap with the absorption spectra of LHC, either the chlorosome or FMO/baseplate part, which suggests that the pigments absorb some amount of light. While the absolute intensities of LEDs are sufficiently different, it should be noted that the photon flux densities in the range of the LHC absorption for 700 nm LED, see **Fig 4(a)**, and 850 nm LED with a band-pass filter, inset in **Fig 4(f)**, are comparable.

Assuming that OD at 625 nm is proportional to the biomass of the culture, the normalized absorption spectra, **Fig. 4**, can be used to compare the amounts of pigment produced by bacteria grown in different conditions. The total area of the peak characterizes the amount of pigments while the ratio of 750 nm and 800 nm peaks provides information about relative concentration of BChl *c*/BChl *a* in LHCs. The latter statement is based on the assumptions that the absorption cross sections of different BChls are almost the same and most of BChl *a* molecules are concentrated in FMO protein complexes. The cultures grown with 750 nm LEDs have largest overlap of the emission spectra with the absorption spectra of LHCs, thus we use it as a reference.

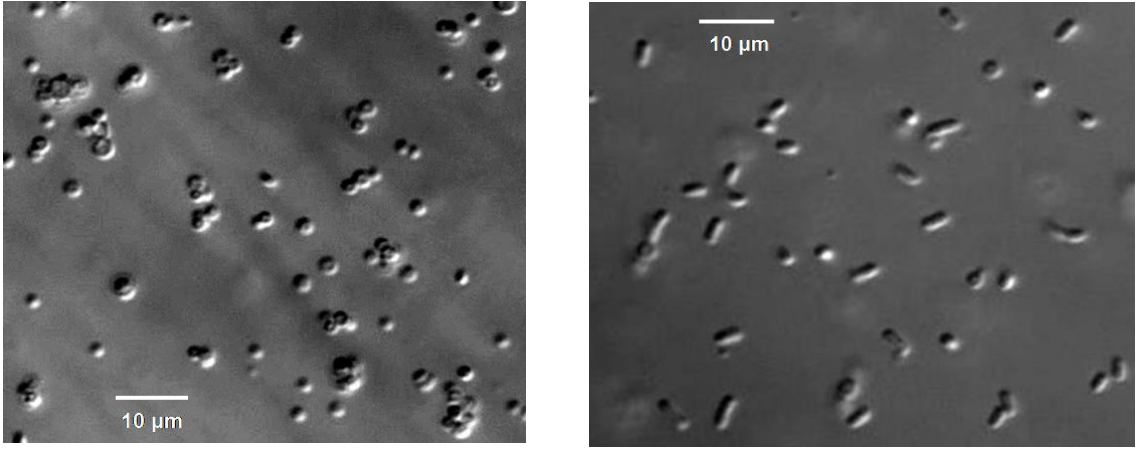

**Figure 3.** Light microscope images of *Cba. tepidum* cultures irradiated with (a) 750 nm LEDs and (b) 850 nm LEDs with a band-pass filter.

The histogram in **Fig. 5(a)** shows that cultures grown with 850 nm LEDs, both with and without the band-pass filter, have substantially more pigments per unit of mass as compared to the culture grown with "optimal" 750 nm LEDs. In contrast, the culture grown with 700 nm LEDs have almost the same amount of pigments as 750 nm culture. The 700 nm cultures also produce the chlorosome with a red-shifted $Q_y$-band maximum, see **Fig. 5(b)**. Previously, this shift was associated with methylation of BChl *c* molecules as a reaction to low-light intensities (10,13). For 850 nm and 850 nm cultures with band-pass filter a less pronounced average shift of the $Q_y$-band was observed.

To estimate the relative concentration of BChl *c*/BChl *a* the normalized spectra were fitted with two Gaussian functions and the short wavelength tail was neglected. Differences in pigment aggregation in the chlorosome and other structural units may slightly affect these estimates. To verify that we additionally performed a set of measurements where pigments were extracted from the cells and extinction coefficients of BChls were measured. The details of the latter measurements are described in Materials and Methods section.

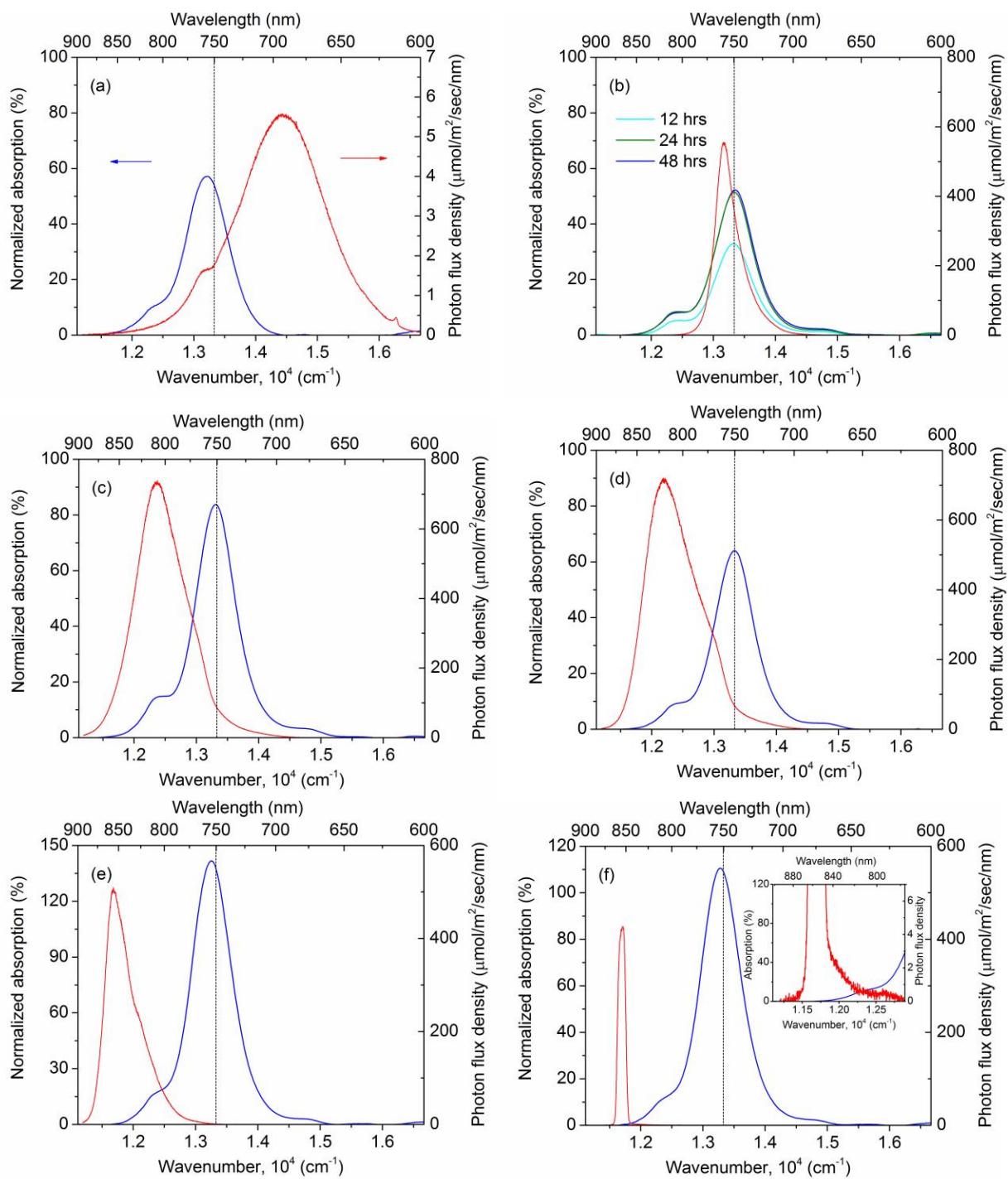

**Figure 4.** Absorption spectra of bacterial cultures grown with different LEDs, blue line, together with the emission spectra of LEDs, red line. Panels (a)-(f) correspond to cultures grown with 700, 750, 780, 800, 850 nm LEDs and also 850 nm LEDs with a band-pass filter. The absorption spectra were taken after the peak intensity saturated, the baseline was removed and the peak intensity was normalized to the absorption at 625 nm. The photon flux density is calculated per 1 nm of the wavelength. The inset in panel (f) shows the overlap of the LED emission spectrum with the absorption peak. Additionally, panel (b) illustrates the time dynamics of the absorption peak.

The BChl *c* molecules practically fill the entire volume of the chlorosome. Thus, the amount of pigments scales as $N_{\mathrm{BChl}c} \propto R^3$. In contrast, the BChl *a* pigments are concentrated on the surface of the chlorosome, and their amount scales as $N_{\mathrm{BChl}a} \propto R^2$. If the absorption cross section is proportional the amount of pigment, then the ratio of the peak areas is proportional to the radius of the chlorosome, $N_{\mathrm{BChl}c} / N_{\mathrm{BChl}a} \propto R$. Thus, the variation in the radius can be monitored by the difference in the peak ratio. We can deduce from **Fig. 5(c)** that the average radius of the chlorosomes in cultures grown with 850 nm LEDs is about 30% larger as compared to 750 nm cultures. It should be noted that the larger size of chlorosomes in 850 nm cultures can solely account for the increase in the pigment amount given in **Fig. 5(a)**. However, in our opinion, the dynamics is more complex. For instance, as it is shown in **Fig. 4(b)** for the 750 nm culture, the chlorosome peak grows about 2 times on the timescale from 12 to 24 hours. At the same time, the relative intensities of the BChl *a* and BChl *c* peaks remain almost the same for this case. Finally, it should be also noted that the dependence of the chlorosome size in green sulfur bacteria on the light intensity, but not the spectral distribution, was reported previously (13, 25).

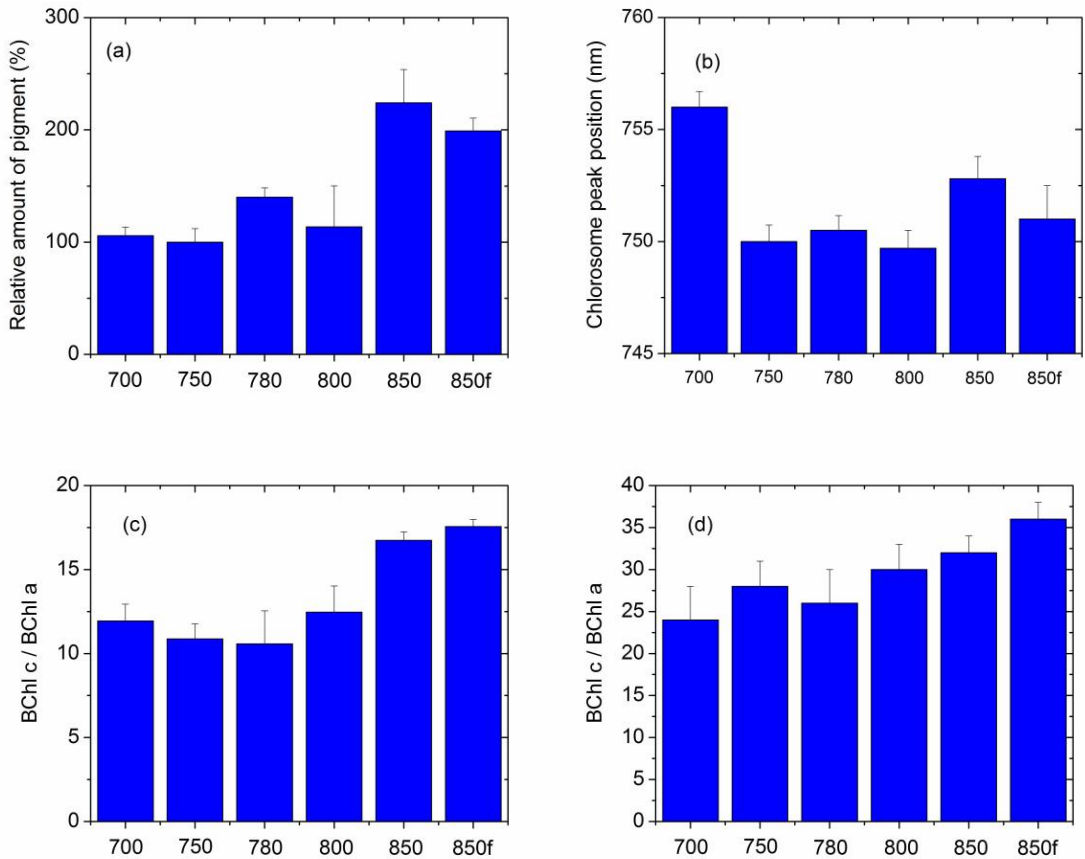

**Figure 5.** (a) Relative amount of pigment in different cultures as compared to the culture grown with 750 nm LED; (b) BChl *c* peak position; (c) relative amount of BChl *c* / BChl *a* estimated from the absorption spectra of the cultures. (d) relative amount of BChl *c* / BChl *a* estimated from the absorption spectra of pigments extracted from the cultures. The error bar corresponds to standard error of mean for 3 sets of experiments with 6 biological replications each.

## DISCUSSION

In order to characterize the obtained growth rates and compare them with the data available from literature we applied the following model. We assumed that the cuvettes with the bacterial cultures are irradiated homogeneously. Then, we estimated the growth rate as a function of absorbed light energy using the relation introduced previously by Baly (26) and Tamiya (27),

$$r = \frac{\alpha r_s I}{r_s + \alpha I}. \tag{2}$$

For low light intensities, $I$, Eq. (2) provides a linear dependence of the growth rate with a slope $\alpha$. The rate saturates to the value $r_s$ at high intensities. While various models were introduced for photosynthetic organisms (28) all of them show a qualitatively similar rate dependence. To account for the spectral profiles of LEDs we weighted the photon flux densities $P(\omega)$, see **Fig. 4**, with the normalized absorption spectra of the cultures $A(\omega)$ as

$$I = \int A(\omega) P(\omega) d\omega. \tag{3}$$

This weighted intensity characterizes the amount of light energy that can be absorbed by LHCs. The resulted growth rates together with the data from Refs. (13-15, 24) are shown in **Fig. 6**.

For the high and intermediate light intensities, our results are comparable with the data from the other studies. In contrast, the estimated low-light intensity growth rate is about 5 times larger than the previously reported growth rates. This discrepancy cannot be described by accounting for the spatially inhomogeneous irradiation of the culture (the 850 nm bacterial cultures where irradiated by single LEDs which provided inhomogeneous light intensity on the culture) in our model.

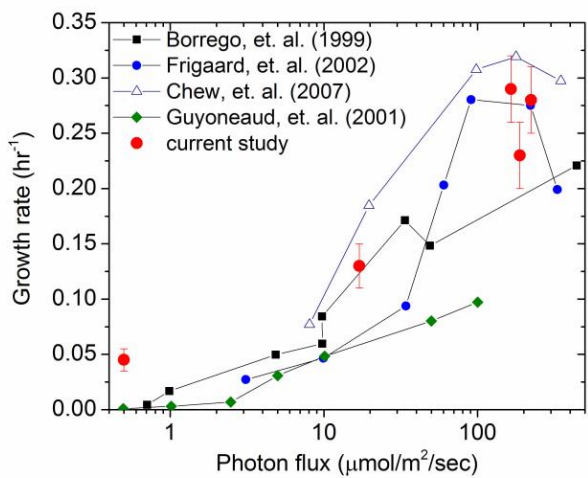

**Figure 6.** Estimated growth rate as a function of the photon flux. The data is compared to previous measurements from Refs. (13-15) for *Cba. tepidum* cultures and Ref. (24) for *Prosthecochloris aestuarii*.

The dependence of the growth rate on the light intensity is either linear or sub-linear. Thus, our estimates should provide a lower bound for the growth rate. One of the possibilities which we consider is that in white or fluorescent lights used in previously reported experiments, the fraction of the spectrum overlapping with the $Q_y$-band of LHC may be smaller as compared to that of a flat emission spectrum. Those spectra are, usually, shifted towards shorter wavelengths. While for high intensities close to saturation of the growth rate the cultures are less sensitive to the variation in energy flux, the effect is more important for low intensities.

Another interesting finding is that the weighted photon flux for 700 nm cultures was about $I \approx 1.5$ µmol/m$^2$/sec, which is twice as large as the weighted flux for the cultures illuminated by 850 nm LEDs when employing a filter. Nevertheless, the former cultures were not able to grow well. This suggests that a model with a 100% energy transfer efficiency of LHCs may be oversimplified for a quantitative characterization of these experiments. For example, in 700 nm cultures some amount of absorbed energy can be lost during the transfer through the LHC due to the exciton trapping and recombination. However, to our knowledge neither experimental pump-probe studies of energy transfer (29-31) in chlorosomes nor microscopic computational models (32,33) support this hypothesis. We believe that more extensive studies are required in order to verify this observation.

We observed the red-shift of the chlorosome $Q_y$-band for the most of low-intensity experiments. In fact, the largest shift is obtained for 700 nm cultures. In this case the chlorosome band is shifted away from the spectrum of available light, which reduces the amount of absorbed energy (energy absorption is controlled by the overlap of these two spectra). This suggests that green sulfur bacteria do not maximize the absorption cross section of LHC by tuning its frequency. Instead, the red-shift of the chlorosome band results in a larger spectral overlap between the chlorosome band and the absorption bands of the baseplate and FMO, thus increasing the energy transfer efficiency between the structural subunits of LHC.

The grown bacterial cultures can be classified in two groups: the group of high energy flux (750 nm, 780 nm and 800 nm) and that of a low energy flux (850 nm with and without filter). These groups differ in: (a) the chlorosome peak position, (b) ratio of BChl *a* and BChl *c* peaks, and (c) amount of pigment per unit of mass, see **Fig. 5**. The differentiation in LHCs occurs on a timescale shorter than 12 hrs, as discussed in the SI. Thus, provided that the mutation rate of bacteria is rather slow (34) the modifications of the LHC complexes can be assigned to bacterial acclimation rather than adaptation, which should take much more time. The observed timescale is actually shorter that the acclimation timescale of cyanobacteria (35).

Finally, we would like to draw attention to the recent report where green sulfur bacteria were found near thermal vents on the bottom of the Pacific Ocean (2). The radiative emission from thermal vents can be described by a blackbody radiation model with a corresponding temperature of several hundred degrees Celsius (36). Our simple estimates give that the photon flux density at the FMO absorption peak, 810 nm, should be about one order of magnitude stronger than the flux at 750 nm, where BChl *c* pigments absorb, which almost compensates the difference in the absorption cross sections of these units, **Fig. 5(c)**. Thus, one can hypothesize that the low-frequency light absorption may be used by green sulfur bacteria in order to survive in natural habitats.

CONCLUSIONS

In this study, we analyzed the response of *Chlorobaculum* [*Cba.*] *tepidum* species of green sulfur bacteria to light with constrained spectral properties. Our results suggest that *Cba. tepidum* tune up the light absorption properties mostly by: (a) increasing number of chlorosomes per unit of mass, and (b) changing the size of the chlorosomes. We find that the red shift of the chlorosome absorption peak, previously reported as a response to low-light conditions, may result in reduced absorption efficiency. All the reported changes occur on the timescale comparable to the bacteria doubling time and can be considered as acclimation. Finally, we observe that the cultures can grow with 850 nm light, which has a negligible overlap with the chlorosome $Q_y$-band and is, possibly, absorbed by Fenna-Matthews-Olson protein complexes and reaction centers. This result supports the recent finding that green sulfur bacteria can live on the bottom of the ocean where only geothermal lights are available.

MATERIALS ANS METHODS

*Chlorobaculum [Cba.] tepidum* TLS strain was cultured as reported in Refs. (22,37). All *Cba. tepidum* TLS cultures reported in this work were grown anaerobically at $T = 47 \pm 3$ ºC, and cell growth was estimated turbidimetrically at the 625 nm wavelength (23). The growth media used in this report, with $NaHCO_3$ and $NH_4Cl$ included as the carbon and nitrogen sources, were described previously (11). All chemicals were purchased from Sigma-Aldrich. 2% culture (50-fold dilution) in the late exponential growth phase was used to inoculate fresh media. For cultures illuminated with 700, 750, 780, 800 and 850 nm LEDs, 6 biological replicates were performed with each wavelength of LED. Additionally, the experiments were repeated 3 times with the time interval of several months.

700 nm light-emitting diodes were purchased from AND OPTOELECTRONICS, 750 nm, 780 nm and 800 nm LEDs where obtained from Marubeni America Corporation, 850 nm LEDs from OSRAM and 940 nm QEC113 LED from Fairchild Semiconductor Inc. The 850 nm band-pass filter with a 25 mm diameter, 85% transmission, and FWHM (full width at half maximum) spectral width $\Delta\lambda \sim 10$ nm was purchased from Edmund Optics Inc. The irradiation intensity of LEDs was measured using a Thorlabs S120C photodiode sensor (Thorlabs, Inc.) and the emission spectra of LEDs were measured using an Ocean Optics USB 4000 spectrometer (Ocean Optics, Inc.). Cell images were taken using a Nikon Eclipse E600 fluorescence microscope (Nikon Instrument Inc.) with cells in the late exponential growth phase.

To determine relative concentrations of BChl *a* and BChl *c* in the cultures 10 mL steady-state cultures illuminated with 700, 750, 780, 800 and 850 nm LEDs were harvested. BChl *a* and BChl *c* were extracted with methanol from cell pellets. The concentrations of BChl *a* and BChl *c* were estimated using the extinction coefficient of BChl *a* in methanol, 54.8 mM$^{-1}$cm$^{-1}$ at 771 nm, and of BChl *c* in methanol, 69.7 mM$^{-1}$cm$^{-1}$ at 667 nm (38).

ACKNOWLEDGEMENT

We thank Dr. Joel Kralj for assisting with the measurement of irradiation intensity and emission spectra of LEDs. We also appreciate discussions of the results with Dr. Alexander Eisfeld and Dr. Dmitrij Rappoport. J.H and A.A.-G. acknowledge support from the Center for Excitonics, an Energy Frontier Research Center funded by the US Department of Energy, Office of Science and Office of Basic Energy Sciences under award DESC0001088. A.A.-G. and S.K.S. acknowledge Defense Threat Reduction Agency grant HDTRA1-10-1-0046. J.K.T. is supported by start-up funds and faculty development fund from Clark University.

# SUPPLEMENTARY INFORMATION

**Figure S1. Experimental set-up for reported LEDs studies.** (a) An arrays of 780 nm LEDs, similar arrays were used for 700, 750, 800 and 940 nm LEDs; (b) 850 nm LED.

**(a)**

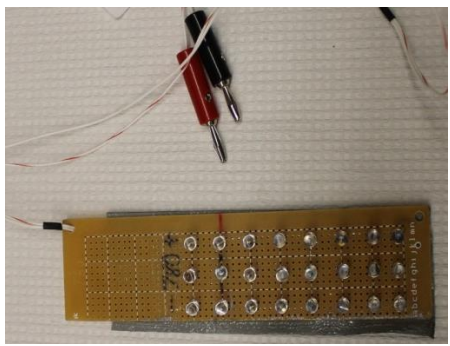

**(b)**

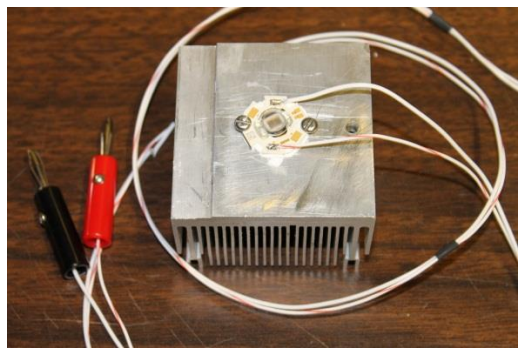

**Figure S2. Bacterial cultures grown with different LEDs.** The cultures grown to saturated density with (a) 750, 780, 800, and 850 nm LEDs, and (b) 700 nm LEDs.

**(a)**

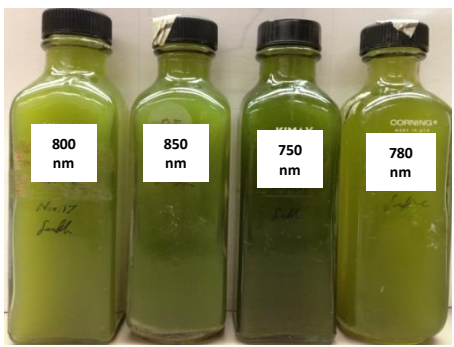

**(b)**

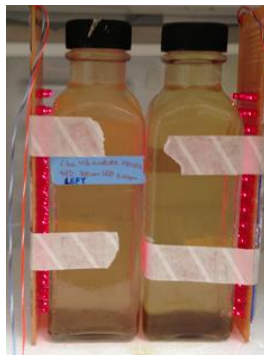

**Figure S3. Population dynamics of bacterial cultures.** The cultures are grown with: (a) 750 nm, (b) 780 nm, (c) 800 nm, (d) 850 nm LEDs, and (e) 850 nm LEDs with a band-pass filter. The blue dots are for data points are taken from 3 different cultures (6 replications each) grown independently with the time interval of several months. The red curves represent sigmoidal fit.

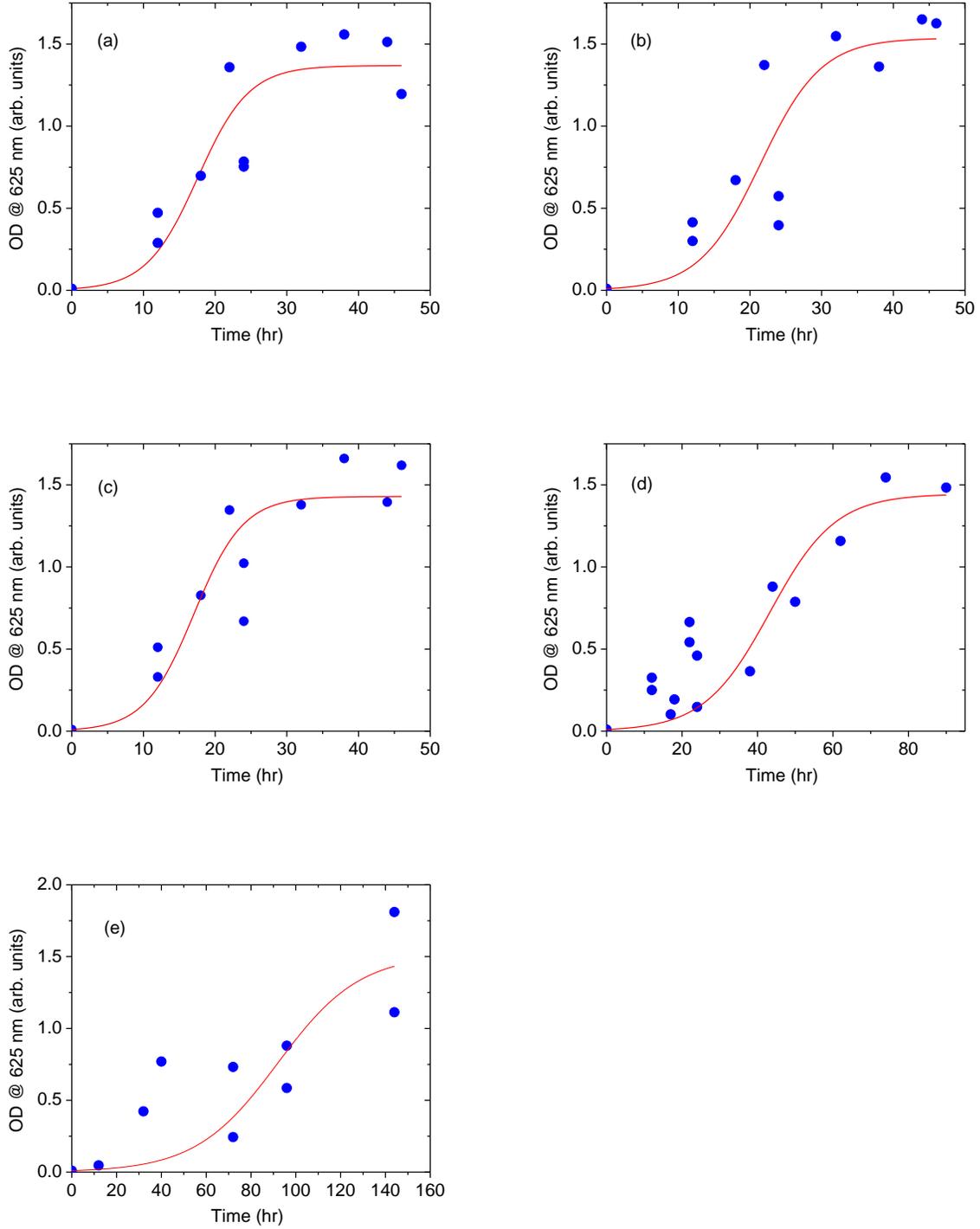

**Figure S4. Time dynamics of LHC complexes.** The grown cultures are differentiated in two groups: high energy flow (750 nm, 780 nm and 800 nm) – blue dots, and low energy flux (850 nm with and without filter) – red dots. (a) Peak area; (b) peak ratio BChl *c*/BChl *a*; and (c) BChl *c* peak position as a function of time.

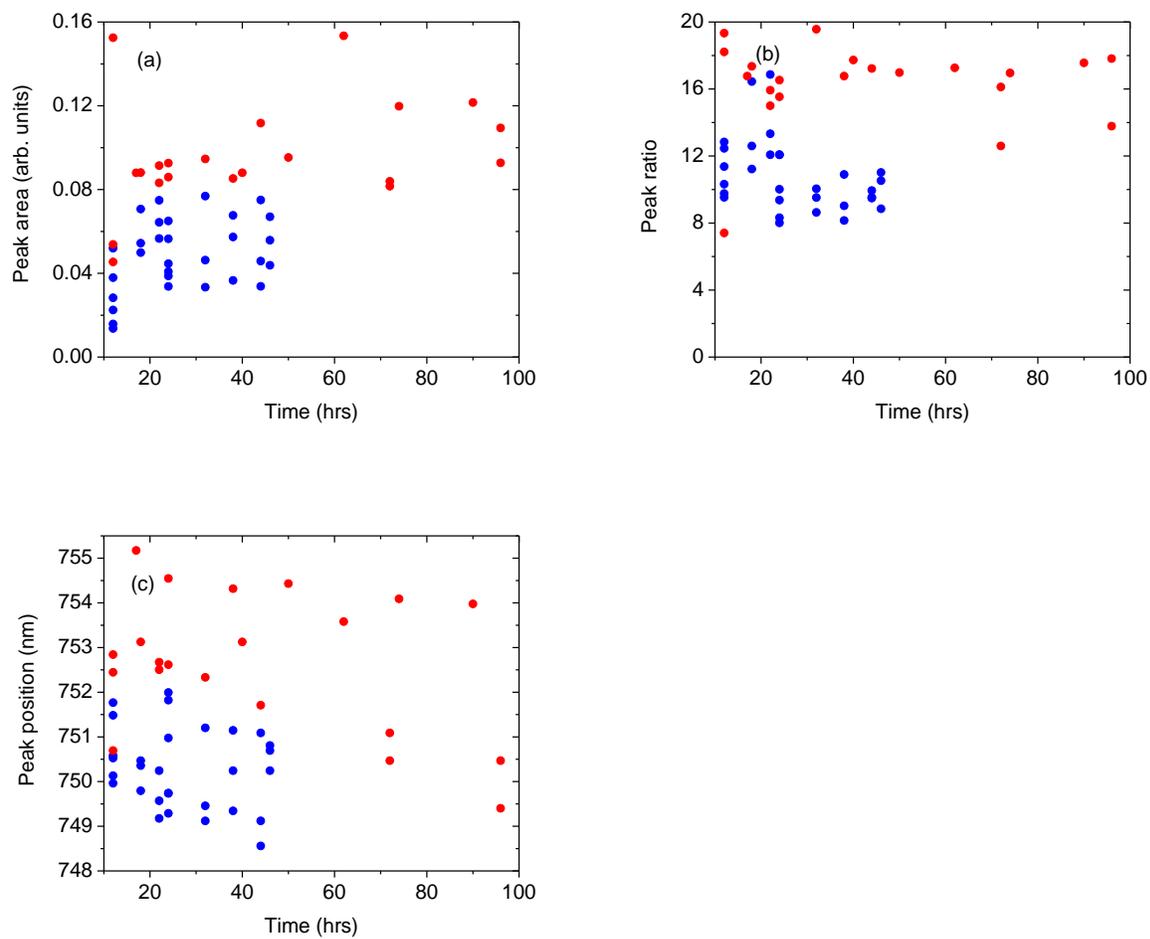